\documentclass[useAMS,usenatbib,fleqn]{mn2e}

\usepackage{graphicx}
\usepackage{float}
\usepackage{amsmath}
\usepackage{amssymb}
\usepackage[labelformat=empty,labelsep=none]{subcaption}
\usepackage{epsfig}
\usepackage{times}
\usepackage{multirow}
\usepackage{color}
\usepackage{gensymb}

\def\gtsim {\gtrsim}   
\def\ltsim {\lesssim}   

\newcommand{\dummytitle}[1]{}

\newcommand{\msun}{{\,\rm M}_\odot }
\newcommand{\atd}{ATLAS$^{\rm 3D}$ }

\defcitealias{pt14}{TK14}
\defcitealias{pt15a}{TK15a}
\defcitealias{pt15b}{TK15b}
\defcitealias{pt18a}{Paper I}

\title[Kinematically atypical galaxies]{The metallicity and elemental abundance maps of kinematically atypical galaxies for constraining minor merger and accretion histories}
\author[P.~Taylor, C.~Kobayashi, and C.~Federrath]{Philip~Taylor$^{1,2}$\thanks{E-mail: philip.1.taylor@anu.edu.au}, Chiaki~Kobayashi$^3$, and Christoph~Federrath$^1$\\
$^1$Research School of Astronomy and Astrophysics, Australian National University, Canberra, ACT 2611, Australia\\
$^2$ARC Centre of Excellence for All Sky Astrophysics in 3 Dimensions (ASTRO 3D), Australia\\
$^3$Centre for Astrophysics Research, School of Physics, Astronomy and Mathematics, University of Hertfordshire, Hatfield, AL10 9AB, UK}

\begin{document}
				
\date{Accepted  Received ; in original form}

\pagerange{\pageref{firstpage}--\pageref{lastpage}} \pubyear{}

\maketitle

\label{firstpage}

\begin{abstract}
Explaining the internal distribution and motions of stars and gas in galaxies is a key aspect in understanding their evolution.
In previous work we identified five well resolved galaxies with atypical kinematics from a cosmological simulation; two had kinematically distinct cores (KDCs), and three had counter-rotating gas and stars (CRGD).
In this paper, we show that i) the KDC galaxies have flattening of stellar [O/Fe] at large galacto-centric radii due to the minor mergers that gave rise to the KDCs, and ii) the CRGD galaxies have an abrupt transition in the gas metallicity maps, from high metallicity in the centre to very low metallicity further out.
These galaxies are embedded in dark matter filaments where there is a ready supply of near-pristine gas to cause this effect.
The non-linear increase in gas metallicity is also seen in the radial profiles, but when the metallicity gradients are measured, the difference is buried in the scatter of the relation.
We also find that all five galaxies are fairly compact, with small effective radii given their stellar masses.
This is because they have not experienced major mergers that kinematically heat the stars, and would have destroyed their unusual kinematics.
In order to detect these signatures of minor mergers or accretion, the galaxy scaling relations or radial metallicity profiles are not enough, and it is necessary to obtain the 2D maps with integral field spectroscopy observations.
\end{abstract}



\section{Introduction}
\label{sec:intro}

Galaxies evolve on timescales much longer than there have been astronomers to observe them.
Therefore, we see only a single instant in their evolution.
By compiling catalogues of many galaxies of different types -- that is, at different stages of evolution -- across cosmic time, we can hypothesise how galaxies grow and evolve.
While gross properties, such as mass and morphology, are useful, the internal distribution and motion of stars and gas are vital to understand the history of an individual galaxy, {and dynamical disturbances, which are believed to be formed by mergers, have been used to understand the formation of early-type galaxies \citep[e.g.,][]{kormendy89,schweizer90,schweizer92,bender92}.}

Just as important is the distribution of chemical elements; these provide a fossil record of the galaxy from when they are locked up in star formation, and chemical abundance ratios provide information on the star formation history \citep[e.g.,][]{ck16nat}.
Galactic archaeology surveys such as Gaia-ESO \citep{GAIAESO}, GALAH \citep{GALAH}, and APOGEE \citep{APOGEE} are set to obtain the chemical profiles of $\sim1$ million stars in the Milky Way, offering unprecedented insight into the history of our Galaxy.
In more distant galaxies, such measurements of individual stars are not possible, but long-slit or integral-field spectroscopy (IFS) can provide {the spatial distribution and radial profiles of gas and stellar metallicity} \citep[e.g.,][]{faber73,davies93,spolaor09,spolaor10,kuntschner10}.
The gradient of these metallicity profiles provides information on the merging history and previous activity of the active galactic nucleus (AGN) of the galaxy \citep[e.g.,][]{ck04,pt17b}.

In recent years, IFS has superseded long-slit spectroscopy as the standard tool for analysing the internal structure of galaxies {in the local Universe}.
Several surveys of hundreds or thousands of galaxies now exist, including SAURON \citep{dezeeuw02}, \atd \citep{cappellari11}, CALIFA \citep{sanchez12}, SAMI \citep{croom12,green18}, SLUGGS \citep{brodie14}, MASSIVE \citep{ma14}, S7 \citep{dopita15,thomas17}, and MaNGA \citep{yan16}, with the next generation of IFS surveys such as Hector \citep{Hector} set to observe 100,000 galaxies.
By fitting single stellar population (SSP) models convolved with a line of sight velocity distribution to every spectrum, maps of line of sight velocity, velocity dispersion, and SSP parameters such as metallicity, [$\alpha$/Fe], and age can be produced.
Furthermore, gas emission lines can also be fitted to yield ionised gas kinematics and abundance maps {from nearby to distant galaxies \citep[e.g.,][]{stott14,wisnioski15}}.

In most galaxies with large-scale, ordered motion, the stars and gas are co-planar and orbit in the same direction.
A significant minority, however, are found to have misaligned gas and stellar kinematics due to the accretion of extra-galactic gas \citep[e.g.,][]{davis11,bryant19}.
A still smaller fraction host a kinematically distinct core (KDC); a change with radius of the position angle of the kinematic axis in stellar velocity maps \citep[e.g.,][]{bender92,krajnovic08,krajnovic11}.
There have been several theoretical studies attempting to replicate and explain the formation of KDCs, finding that gas-rich, equal-mass mergers are required to produce a KDC \citep{jesseit07,hoffman10,bois11,khochfar11,naab14}.
Large-scale cosmological simulations have recently been used to investigate galaxy kinematics, but the detailed origin of KDCs has not yet been addressed \citep{penoyre17,schulze18,sande18}.

In \citet[][hereafter \citetalias{pt18a}]{pt18a}, we presented five galaxies from our cosmological simulation \citep{pt15a} that displayed atypical kinematics; three had counter-rotating gas and stars, and two had a KDC.
These KDCs did not form due to major mergers, as in previous theoretical studies.
The KDC formation mechanism can be summarised as follows: i) the galaxy forms in the field with co-rotating gas and stars; ii) the galaxy falls into a dark matter filament, where the relative orientation of the stellar angular momentum and bulk gas flow leads to the formation of a counter-rotating gas disc (CRGD); and iii) a minor merger, with a galaxy on a retrograde orbit compared to the stellar angular momentum of the primary, deposits counter-rotating stars in the outskirts of the primary {(we note that \citealt{bassett17} found a similar result using idealised simulations of disc galaxy mergers with a mass ratio of 1/10)}.
The resulting KDCs were larger than observed, and were not clearly discernible in luminosity-weighted kinematic maps, which are a closer analogue to kinematic maps generated from IFS data than mass-weighted maps.
We suggest that this is a separate class of KDCs, which has not yet been identified in observations.

The aim of this paper is to analyse the maps of stellar populations and gas and stellar chemistry that are comparable to IFU data, and to determine the distinguishing features of these galaxies, beyond their kinematics, that set them apart from the wider galaxy population.
In Section \ref{sec:sims} we describe our simulation and analysis techniques, and introduce the galaxies with atypical kinematics in Section \ref{sec:gals}.
We compare the properties of these galaxies to the full population of simulated galaxies in Section \ref{sec:results} before presenting our conclusions in Section \ref{sec:conc}.

\section{Simulation Setup and Analysis Procedure}
\label{sec:sims}

\subsection{The Simulation}

The simulation used in this paper is a cosmological, chemodynamical simulation, first introduced in \citet{pt15a}.
Our simulation code is based on the smoothed particle hydrodynamics (SPH) code {\sc gadget-3} \citep{springel05gadget}, updated to include:  star formation \citep{ck07}, energy feedback and chemical enrichment from supernovae \citep[SNe II, Ibc, and Ia,][]{ck04,ck09} and hypernovae \citep{ck06,ck11a}, and asymptotic giant branch (AGB) stars \citep{ck11b}; heating from a uniform, evolving UV background \citep{haardt96}; metallicity-dependent radiative gas cooling \citep{sutherland93}; and a model for black hole (BH) formation, growth, and feedback \citep{pt14}, described in more detail below.
We use the initial mass function (IMF) of stars from \citet{kroupa08} in the range $0.01-120\msun$, with an upper mass limit for core-collapse supernovae of $50\msun$.

The initial conditions for the simulation consist of $240^3$ particles of each of gas and dark matter in a periodic, cubic box $25\,h^{-1}$ Mpc on a side, giving spatial and mass resolutions of $1.125\,h^{-1}$ kpc and $M_{\rm DM}=7.3\times10^7\,h^{-1}\msun$, $M_{\rm gas}=1.4\times10^7\,h^{-1}\msun$, respectively.
This resolution is sufficient to spatially resolve structure within massive galaxies.
We employ a WMAP-9 $\Lambda$CDM cosmology \citep{wmap9} with $h=0.7$, $\Omega_{\rm m}=0.28$, $\Omega_\Lambda=0.72$, $\Omega_{\rm b}=0.046$, and $\sigma_8=0.82$.

BHs form from gas particles that are metal-free and denser than a specified critical density, mimicking the most likely formation channels in the early Universe as the remnant of Population {\sc III} stars \citep[e.g.,][]{madau01,bromm02,schneider02} or via direct collapse of a massive gas cloud \citep[e.g.,][]{bromm03,koushiappas04,agarwal12,becerra15,regan16a,hosokawa16}.
The BHs grow through Eddington-limited Bondi-Hoyle gas accretion and mergers.
Two BHs merge if their separation is less than the gravitational softening length and their relative speed is less than the local sound speed.
A fraction of the energy liberated by gas accretion is coupled to neighbouring gas particles in a purely thermal form.
 
In previous works, we have compared the simulation used in this paper with another having the same initial conditions, but without the inclusion of any BH physics.
We showed that our model of AGN feedback leads to simulated galaxies whose properties more closely match those of observed galaxies, both at the present day and high redshift \citep{pt15a,pt16,pt17b}, and quantified the effects of AGN feedback on the host galaxy and its immediate environment \citep{pt15b,pt17a}.

\subsection{Generating Maps}

IFU observations of galaxies provide individual spectra across the face of the galaxy.
From each spectrum, average gas and stellar metallicities and chemical abundances can be extracted, as well as stellar age, and gas and stellar kinematics.
In our simulation, these properties are tracked for each gas and star particle, and we produce maps of these quantities using the following method.

Galaxies are rotated so that the net angular momentum of their stars lies along the $z$-axis (i.e., along the line of sight), in keeping with the similar analysis of \citetalias{pt18a}.
We smooth the properties of individual particles over pixels on a regular grid.
Specifically, the average value of a quantity $P$ in the $j^{\rm th}$ pixel is given by
\begin{equation}\label{eq:average}
	\left<P\right>_j = \frac{\sum_i P_i f_{ij} w_i}{\sum_i f_{ij} w_i},
\end{equation}
{ where the sum is over all particles of interest, typically the gas or stars of a galaxy, $f_{ij}$ denotes the fractional contribution of particle $i$ to pixel $j$, and the weights $w_i$ are either particle mass, or, for closer comparison to observational data, $V$-band luminosity ($L_V$) for stars and star formation rate (SFR) for gas (see \citetalias{pt18a} for details)}.
Note that we evaluate (O/Fe)$_j$ as $<$O$>_j$/$<$Fe$>_j$.
This produces qualitatively similar maps as $<$O/Fe$>_j$, but smooths the contribution of individual particles with extreme values of [O/Fe].

\subsection{The Galaxy Sample}\label{sec:gals}

\begin{table}
\caption{Present-day properties of the galaxies presented in Section \ref{sec:gals}.
Stellar mass, gas mass, BH mass, and effective radius are given.
The kinematic feature seen in each galaxy is also listed; counter-rotating gas disc (CRGD) or kinematically distinct core (KDC).}
	\begin{tabular}[width=0.5\textwidth]{cccccc}
		Galaxy & Kinematics & $\log M_*$ & $\log M_{\rm gas}$ & $\log M_{\rm BH}$ & $R_{\rm e}$ \\
		& & $[\msun]$ & $[\msun]$ & $[\msun]$ & [kpc] \\
		\hline 
		ga0045 & CRGD & 10.8 & 9.8 & 6.2 & 3.2 \\
		ga0064 & CRGD & 10.7 & 9.7 & 5.7 & 2.6 \\
		ga0074 & KDC & 10.6 & 9.9 & 6.1 & 2.3 \\
		ga0091 & KDC & 10.5 & 9.8 & 6.0 & 2.4 \\
		ga0099 & CRGD & 10.5 & 9.9 & 5.9 & 2.0 
	\end{tabular}
\label{tab:gals}
\end{table}

\begin{figure}
	\centering
	\includegraphics[width=0.48\textwidth,keepaspectratio]{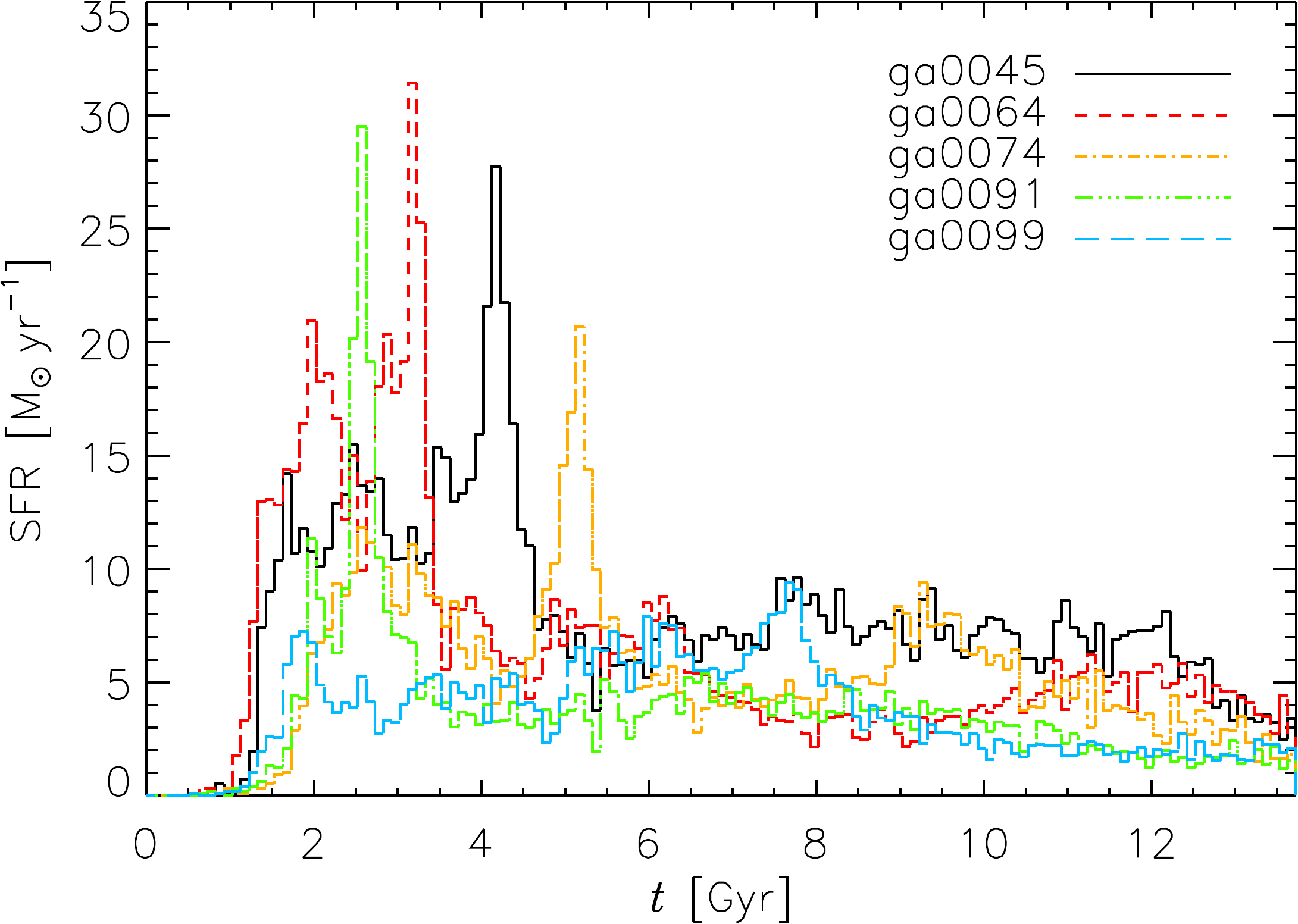}
	\caption{Star formation histories of the five galaxies with atypical kinematics listed in Table \ref{tab:gals}. The time $t$ denotes time since the Big Bang.}
	\label{fig:sfr}
\end{figure}

\begin{figure*}
	\centering
	\includegraphics[width=0.98\textwidth,keepaspectratio]{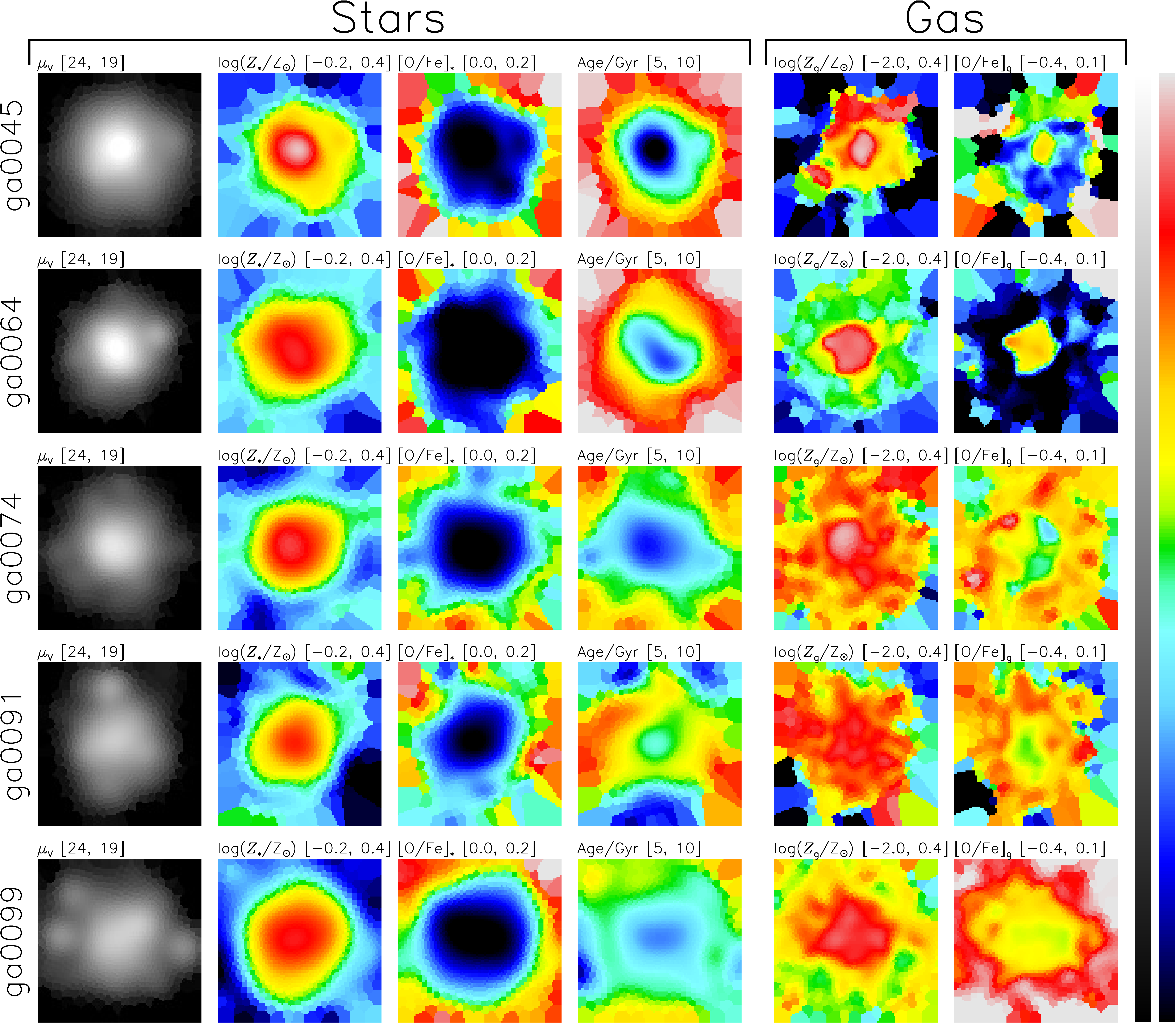}
	\caption{Maps of mass-weighted stellar and gas properties for our five simulated galaxies with atypical kinematics.
	The columns show $V$-band surface brightness (mag arcsec$^{-2}$), stellar metallicity $\log\left(Z_*/{\rm Z}_\odot\right)$, stellar [O/Fe], stellar age, gas metallicity $\log\left(Z_{\rm g}/{\rm Z}_\odot\right)$, and gas [O/Fe], respectively.
	The range of values in each panel is shown in brackets { (low values are blue, high are red)}.
	Each panel is $6R_{\rm e}$ on a side.}
	\label{fig:maps}
\end{figure*}

Galaxies are identified using a parallel Friends-of-Friends (FoF) finder (based on a serial version provided by V.~Springel).
The FoF code associates dark matter particles, separated by at most 0.02 times the mean inter-particle separation, into groups.
Gas, star, and BH particles are then joined to the group of their nearest dark matter neighbour.
Note that the `linking length' of 0.02 used here is smaller than typically adopted in the literature \citep[e.g., 0.2 in ][]{springel05nat,ck07,dolag09,vogelsberger14nat,schaye15}.
In these works, sub-halos are separated from the main FoF groups, whereas we use the smaller linking length to achieve the same result.

Five of the 82 sufficiently well resolved galaxies identified in \citetalias{pt18a} were found to have atypical kinematics; three had counter-rotating gas and stellar components, and two had a KDC.
All of the five galaxies -- denoted ga0045, ga0064, ga0074, ga0091, and ga0099 -- formed in the field, and later fell into dark matter filaments where their unusual kinematic signatures developed through gas accretion and minor mergers.
The basic galaxy properties are summarised in Table \ref{tab:gals}, where the masses are measured for all particles identified by the FoF code.
{The star formation histories of these galaxies are shown in Fig. \ref{fig:sfr}.
All but ga0099 experience a strong peak in star formation at early times, and the formation of a CRGD reduces their star formation rates at late times (see \citetalias{pt18a} for more details).}

\section{Results}\label{sec:results}

The columns of Fig. \ref{fig:maps} show maps of $V$-band surface brightness ($\mu_V$), stellar metallicity $\log\left(Z_*/{\rm Z}_\odot\right)$, stellar [O/Fe], stellar age, gas metallicity $\log\left(Z_{\rm g}/{\rm Z}_\odot\right)$, and gas [O/Fe], respectively for the five galaxies.\footnote{ These maps are not affected by the resolution of our simulation; we reproduced the maps with a random selection of half of the particles and find excellent qualitative and quantitative agreement.}
The maps of stellar populations are qualitatively similar for all of the galaxies.
They have negative metallicity gradients, and most variation in the metallicity maps is radial.
Similarly, [O/Fe]$_*$ varies primarily radially, and all five galaxies have small, positive gradients.
This is because the star formation in the centre { takes place over a longer timescale} than the outskirts, and thus there is more time for SNe Ia to contribute Fe enrichment.
There is slightly more variety in the maps of stellar age, but in general the stars at the centre of the galaxies are younger, on average, than those further out.

The maps of gas properties show greater structure than for stars, and can be separated by kinematic features.
The galaxies ga0045, ga0064, and ga0099 have counter-rotating gas and stars as a result of prolonged accretion of gas from a dark matter filament \citepalias{pt18a}.
These galaxies show abrupt changes in gas metallicity (at $\sim 1.5$, $0.8$, and $1.3R_{\rm e}$, respectively) from the enriched gas towards the centre, to the metal-poor gas recently accreted from the filament (see Section \ref{sec:profiles} for the radial profiles).
Galaxy ga0045 also shows a metal-enhanced region above the centre, which is not caused by a satellite galaxy; from the $\left<v_{xy}\right>$  and $\phi$ maps of \citetalias{pt18a} inflow and outflow gas hit around this area.
By contrast, ga0074 and ga0091, which host a KDC that formed following a minor merger, do not have such an abrupt change in metallicity with radius.
The material accreted from the minor merger orbits away from the centre of the galaxies \citepalias{pt18a}, and the gas of the secondary is already enriched compared to the gas in the filaments, leading to the more extended regions of high metallicity at $\sim 2 R_{\rm e}$.

Among CRGD galaxies, ga0099 has a relatively featureless [O/Fe]$_{\rm g}$ profile with a shallow positive gradient, ga0045 and ga0064 have a region of high [O/Fe]$_{\rm g}$ in its centre.
This is due to a brief but recent ($<200$ Myr) episode of star formation in the centre of the galaxy, from which only core-collapse SNe have contributed to [O/Fe]$_{\rm g}$, raising the value locally.
In KDC galaxies ga0074 and ga0091, [O/Fe]$_{\rm g}$ increases radially from the centre, before dropping again at $\sim 2R_{\rm e}$ (see Section \ref{sec:profiles} for more discussion).

Fig. \ref{fig:maps} shows mass-weighted quantities, whereas observations produce luminosity-weighted quantities.
When we produce the same maps weighting by $L_V$ for stellar quantities and SFR for gas, they are qualitatively and, except for stellar age, quantitatively similar, and the arguments presented above hold.

\subsection{Radial Metallicity and [O/Fe] Profiles}
\label{sec:profiles}

\begin{figure*}
	\centering
	\includegraphics[width=0.98\textwidth,keepaspectratio]{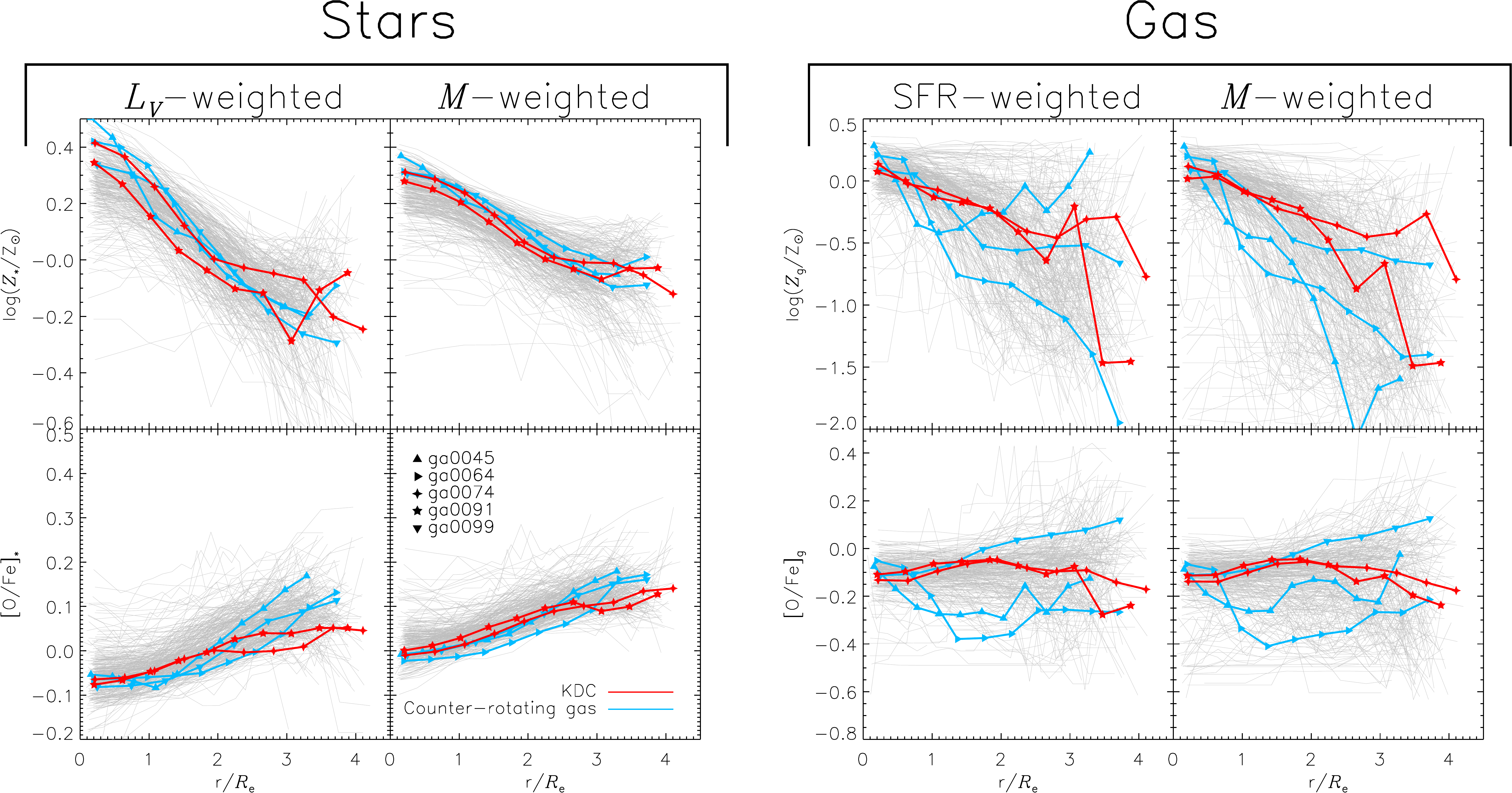}
	\caption{Metallicity (top row) and [O/Fe] profiles (bottom row) as a function of galacto-centric radius (in units of effective radius, $R_{\rm e}$) for the full galaxy population (grey) and those galaxies with atypical kinematics.
	The two leftmost columns are for $L_V$- and mass-weighted stellar profiles, respectively, and the two rightmost are for SFR- and mass-weighted gas quantities.
	Blue triangles are for galaxies with a CRGD, and red { stars} are KDCs.}
	\label{fig:zprofile}
\end{figure*}

We show in Fig. \ref{fig:zprofile} the metallicity (top row) and [O/Fe] (bottom row) profiles for stars (left-hand panels) and gas (right-hand panels).
The full galaxy population is shown in grey, and the galaxies with atypical kinematics have blue triangle symbols for galaxies with a CRGD, and red { stars} for KDCs.
Both mass- and $L_V$-weighted stellar metallicity profiles mirror the full population, and there are no statistically significant distinguishing features.
In general, both stellar metallicity and [O/Fe] gradients become steeper when weighted by luminosity.
This is because Fe-enhanced populations are younger and brighter.
A similar effect is seen for the observed metallically gradients in early-type galaxies \citep{goddard17}.
{ Note that in this work we do not split our simulated galaxies by type, but see Fig. 3 and the associated discussion of \citet{pt17b}.}
 
For all of the KDC galaxies, however, the [O/Fe]$_*$ profiles become shallower at $r\gtsim2.5R_{\rm e}$.
As described in detail in \citetalias{pt18a}, these galaxies acquire a KDC due to a minor merger with a secondary galaxy that deposits its stars at large radii.
The low-mass secondary galaxies formed stars over a longer timescale than the old stellar populations at the outskirts of most galaxies (see Fig. \ref{fig:ageprof}), and had correspondingly more time for enrichment of Fe by SNe Ia, leading to lower average stellar [O/Fe] in the outskirts of the KDC galaxies compared to the full galaxy population.
There is no such distinguishing feature for the stellar metallicity.
Therefore, it is very important to measure elemental abundances at outskirts of galaxies in order to find this type of KCD in observations, which is not possible for stellar populations.

The gas-phase metallicity profiles show greater variation on the whole than the stellar metallicity profiles.
The five galaxies with atypical kinematics (red and blue lines) lie within the envelope of the profiles of the full population (grey).
However, as discussed for Fig. \ref{fig:maps}, the three CRGD galaxies { (blue triangles)} show a rapid increase of $Z_{\rm g}$ towards the centre, and the profiles are not well fit by a straight line, which is also the case for luminosity-weighted $Z_*$ profiles; the slope changes appear at $\sim 1.5$, $0.8$, and $1.3R_{\rm e}$, respectively.
When weighted by SFR, one of the highlighted profiles (ga0045) has an upturn to high metallicity at $\gtsim 2 R_{\rm e}$; this is because star formation is more likely to take place in metal-rich gas that can cool efficiently (see Section 3.3 for more discussion).

The [O/Fe]$_{\rm g}$ profiles are shown in the lower-right panels of Fig. \ref{fig:zprofile}.
The two KDC galaxies (red { stars}) have very similar profiles with a peak at $r\sim 1.5R_{\rm e}$.
Gas-phase [O/Fe]$_{\rm g}$ can vary due to i) recent star formation and O production, ii) delayed Fe enrichment by SNe Ia, and iii) mass-loss from stellar populations having [O/Fe]$_*$.
For these galaxies, we found no evidence for enhanced star formation following the mergers that gave rise to the KDCs in \citetalias{pt18a}, so it is unlikely to have the option i) at $\sim 1.5R_{\rm e}$.
Since [O/Fe]$_{\rm g}$ is roughly the same as [O/Fe]$_*$ there, this gas should mainly come from mass-loss.
$1.5R_{\rm e}$ is also roughly the same as the radius of the KDC (Paper I), which might block SN Ia enriched gas falling in.
It is also clear that two of the CRGD galaxies (ga0045 and ga0064; blue triangles) have profiles that differ significantly from a straight line; these are the galaxies identified in Section \ref{sec:scale} as having unusual values of the [O/Fe]$_{\rm g}$ gradient.
In these galaxies, SN Ia enriched gas falls in near the centre, and the central [O/Fe]$_{\rm g}$ is high because of the option i), as discussed for Fig. \ref{fig:maps}.

\subsection{Radial Age Profiles and Gradients}\label{sec:age}

\begin{figure*}
	\centering
	\includegraphics[width=0.98\textwidth,keepaspectratio]{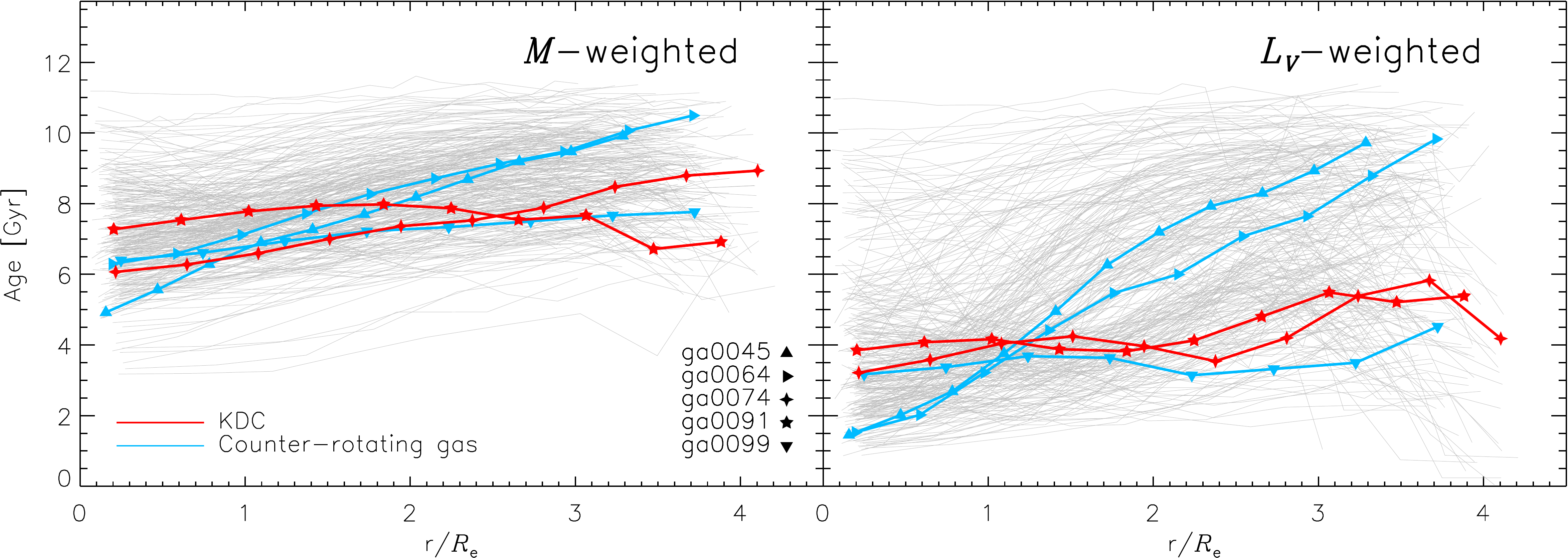}
	\caption{Profiles of average stellar age with galacto-centric radius.
	Grey lines are for the full galaxy population, and red { stars} and blue triangles are for the galaxies with KDCs and counter-rotating gas and stars, respectively.
	The left-hand panel shows mass-weighted ages, and the right-hand $L_V$-weighted ages.}
	\label{fig:ageprof}
\end{figure*}

We estimate ages using equation \eqref{eq:average} with $P={\rm age}$ and $w=L_V$ or mass; this is directly equivalent to the method used in observations where linear combinations of SSP models are used to reproduce the observed spectrum in each spaxel, and the average age of that spaxel is then the weighted mean age of the SSP models \citep[e.g.,][]{zheng17}.
In order to produce radial profiles from the maps of Fig. \ref{fig:maps}, we treat each Voronoi cell as a single point located at the generating point of that cell.
Finally, we bin the data by galacto-centric distance, and adopt the median value of each bin; the resulting age profiles are shown in Fig. \ref{fig:ageprof}.

In the left-hand panel we show mass-weighted age profiles as a function of radius, expressed in units of $R_{\rm e}$, and in the right-hand panel are $L_V$-weighted age profiles.
$L_V$ traces young stellar populations, causing many of the profiles in the right-hand panel to have lower ages, especially at small radii.
In either weighting scheme, there are no distinguishing features in the age profiles of the kinematically interesting galaxies that would set them apart from the full population.

\begin{figure}
	\centering
	\includegraphics[width=0.48\textwidth,keepaspectratio]{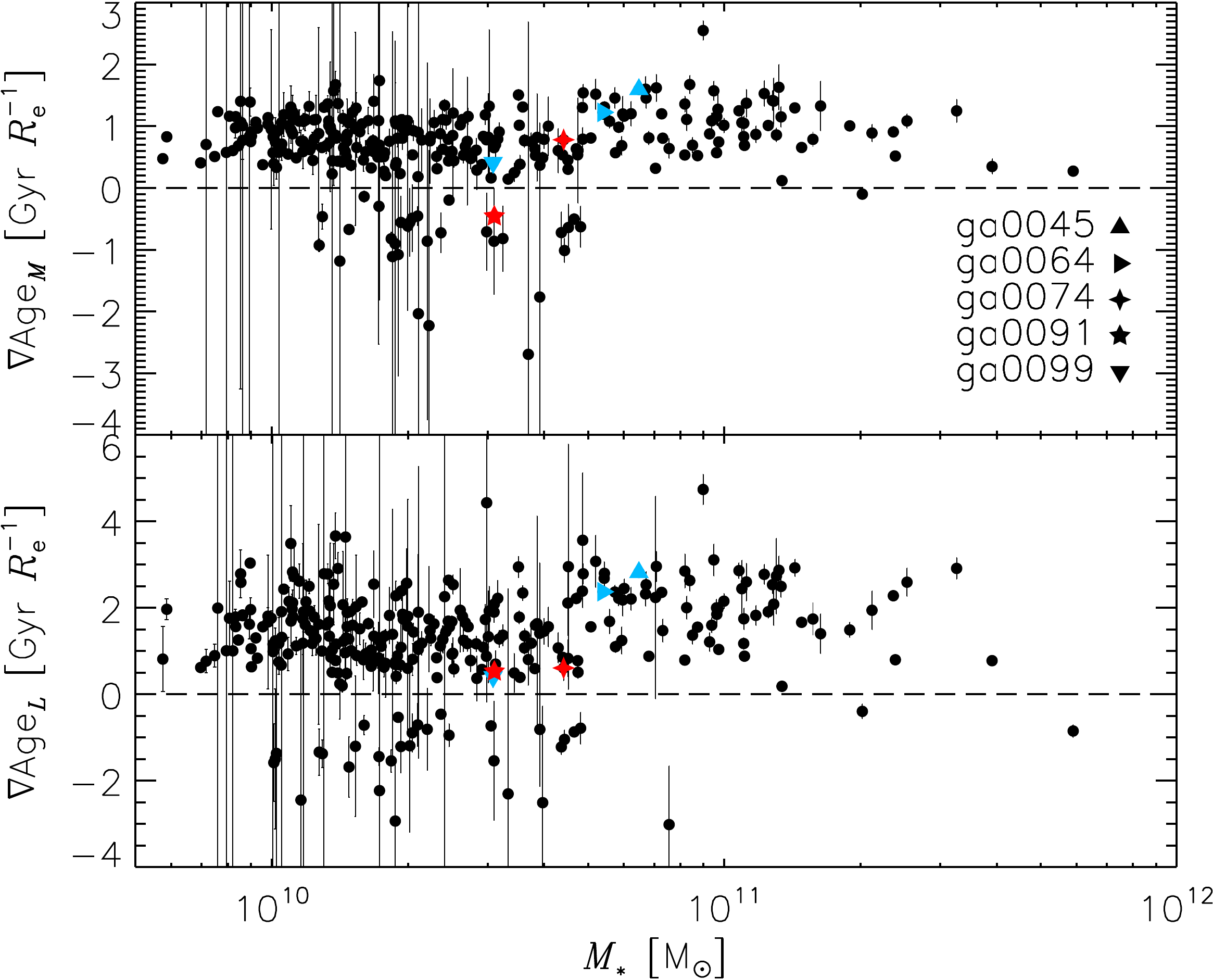}
	\caption{Age gradients as a function of stellar mass.
		The top panel is for mass-weighted ages, the lower for $L_V$-weighted.
		Red { star} and blue triangle symbols show galaxies with KDCs and CRGDs, respectively.}
	\label{fig:agegrad}
\end{figure}

From the profiles of Fig. \ref{fig:ageprof} we can derive age gradients, $\nabla {\rm Age}$,  for our galaxies.
Throughout this paper, we use $\nabla$ to denote $\frac{{\rm d}}{{\rm d}\left(r/R_{\rm e}\right)}$.
We fit a linear function of the form ${\rm Age} = {\rm Age}_0 + \nabla {\rm Age}\times r/R_{\rm e}$ to each profile, and estimate errors on the fitted parameters using a bootstrapping technique.
The gradients of the mass- and $L_V$-weighted age profiles are shown in the top and bottom panels of Fig. \ref{fig:agegrad}, respectively.
Most galaxies have {small} positive age gradients with mass-weighted values $\sim1$ Gyr\,$R_{\rm e}^{-1}$, except the most massive galaxies where the gradient is closer to 0.
Some intermediate-mass galaxies ($10^{10}\ltsim M_*/\msun\ltsim10^{11}$) have negative gradients, indicating that star formation occurred most recently away from the centre of the galaxies.
In these galaxies, the black hole has grown sufficiently massive that AGN feedback is strong enough to quench central star formation \citep{pt17a}, leading to an older central stellar population and negative age gradients.
In more massive galaxies, star formation was quenched earlier, and an old stellar population exists throughout these galaxies, giving rise to shallower gradients.
This is likely to be exacerbated by the effect of major mergers mixing stars of different ages, as also happens to the metallicity gradients of massive galaxies \citep{ck04,pt17b}.
$L_V$-weighted age gradients are qualitatively similar, but tend to be larger in magnitude for both positive and negative gradients.
$L_V$ traces young stars, and the steeper gradients are due to the fact that recent star formation has not taken place uniformly across the galaxy.
This effect is also seen in observations { of both early- and late-type} galaxies \citep{goddard17}.

The blue triangle and red { star} symbols in Fig. \ref{fig:agegrad} show the galaxies with CRGDs and KDCs, respectively.
These galaxies do not have preferentially steep or shallow gradients, and {KDC galaxies tend to have flatter gradients (luminosity weighted in particular), though the scatter is large in this mass range}.
The atypical kinematics seen in these galaxies form at late times \citepalias{pt18a}, and the galaxies do not experience major mergers or late bursts of star formation.
This means that the processes that gave rise to the age gradients in these galaxies likely happened before the CRGDs and KDCs formed.
Although it is likely that the radial age profile of galaxies hosting a KDC changes when the KDC forms, due to the accretion of stars from the secondary galaxy, this effect will depend on the age of the accreted galaxy and cannot be used to help identify a galaxy with a KDC.

\subsection{Scaling Relations}
\label{sec:scale}
\begin{figure*}
	\centering
	\includegraphics[width=0.98\textwidth,keepaspectratio]{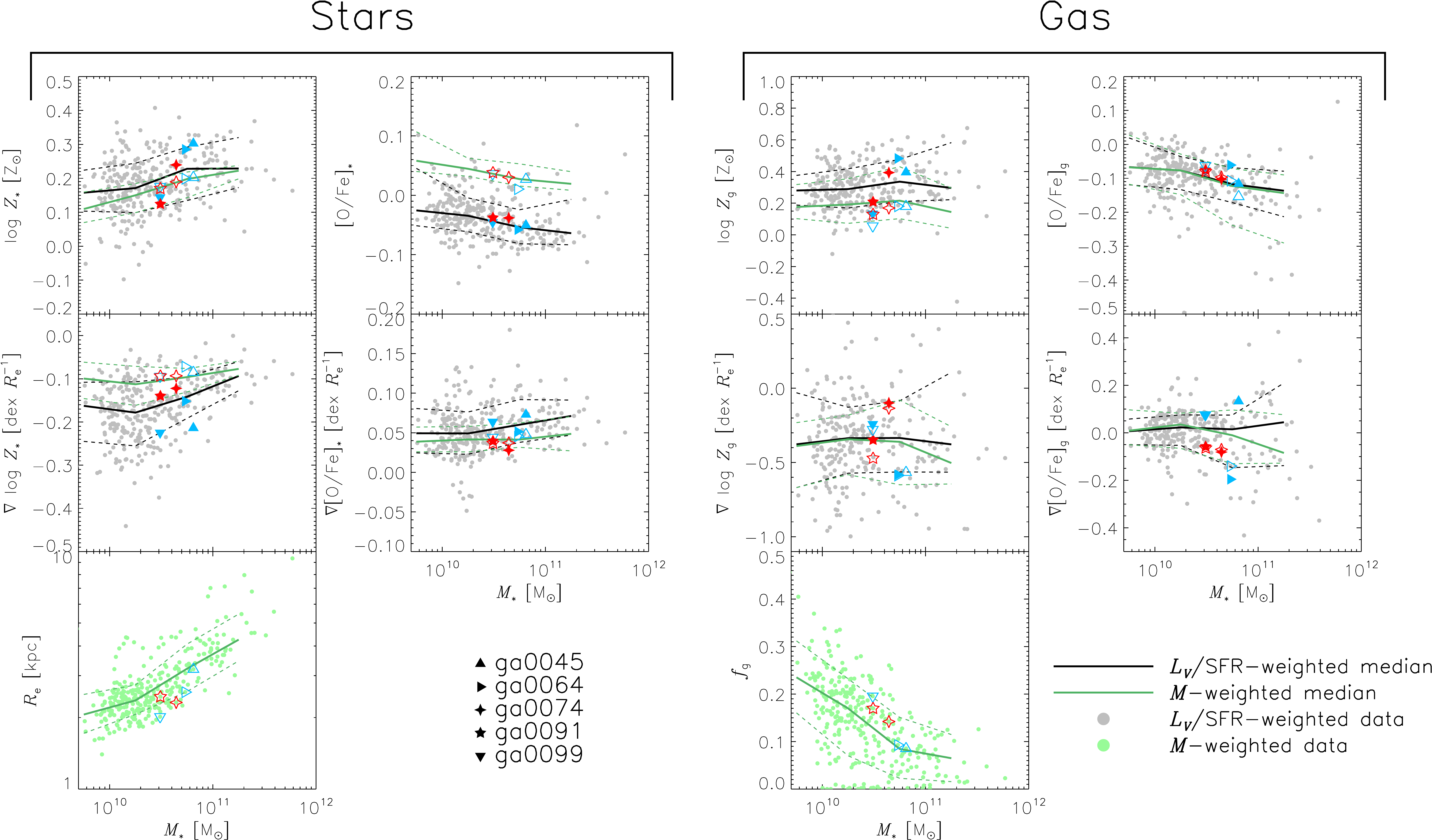}
	\caption{Scaling relations for our simulated galaxies.
	Metallicity and [O/Fe], and the gradients of these quantities, are shown for both stars (left columns) and gas (right column) as functions of stellar mass.
	Effective radius, $R_{\rm e}$, and gas fraction, $f_{\rm g}$, are also shown on the bottom row.
	Faded black points show quantities weighted by $L_V$ (for stars) or SFR (for gas), and faded green points are for mass-weighted quantities.
	Solid black and green lines denote the median light- or mass-weighted relations, and dashed lines show $16^{\rm th}$ and $84^{\rm th}$ percentiles.
	Blue triangles are for the galaxies with a CRGD, and red { stars} are for KDCs; { filled stars and triangles are for $L_V$- or SFR-weighted quantities, open symbols are mass-weighted}.}
	\label{fig:scale}
\end{figure*}

Fig. \ref{fig:scale} shows stellar (left columns) and gas-phase (right columns) scaling relations for our population of simulated galaxies, as well as the two KDC galaxies (red { stars}) and the three CRGD galaxies (blue triangles).
These datasets have been described in detail before (mass-metallicity relations and $R_{\rm e}$-$M_*$ relation in \citealt{pt15a,pt16}, and metallicity and {elemental abundance ratio radial }gradients in \citealt{pt17b}).
Note, however, that these stellar metallicity and abundance gradients shown in this paper are with respect to linear distance (expressed in units of $R_{\rm e}$), rather than $\log$ distance as in \citet{pt17b}.
Nevertheless, the results are qualitatively similar, and the minimum in $\nabla \log Z_*$ at $\sim10^{10}\msun$ is visible in both datasets \citep{spolaor09,spolaor10,kuntschner10,pt17b}.
This mass is consistent with the transition mass for star formation in \citet{pt17a}.
Here we are concerned with the position of the five kinematically atypical galaxies within the distributions; these are shown by the blue triangle symbols (CRGDs) and red { stars} (KDCs).
$L_V$- or SFR-weighted values are shown by faded black points, and the median of these and mass-weighted relations are shown by the black and green lines, respectively.

The stellar metallicity, [O/Fe], and their gradients, of the kinematically atypical galaxies are near the median relations of the full population, which is expected from the fact that these galaxies do not undergo major mergers.
The stellar mass--metallicity relation is set during the peak of star formation \citep[$z\sim2$,][]{pt16}, and gradients are strongly affected by major mergers \citep{pt17b}.
Although the five galaxies of note lie very close to the median [O/Fe]$_*$--$M_*$ relation, we remark that the difference in the $L_V$- and mass-weighted median relationships is due to the fact that the young stars traced by $L_V$ formed from gas that had more time to be enriched by Fe from SNe Ia than on average, causing the distribution to shift down.
In contrast, the effective radii ($R_{\rm e}$) of these galaxies are smaller than average, given their masses (bottom panels).
This is because the galaxies do not undergo major mergers that can kinematically heat the stars, which is also required to allow the atypical kinematic features to survive to the present day.
The $R_{\rm e}$ dependence on the merging history is found in \citet{ck05}, who suggested that this is the origin of the scatter in the fundamental plane.

\begin{figure}
	\centering
	\includegraphics[width=0.48\textwidth,keepaspectratio]{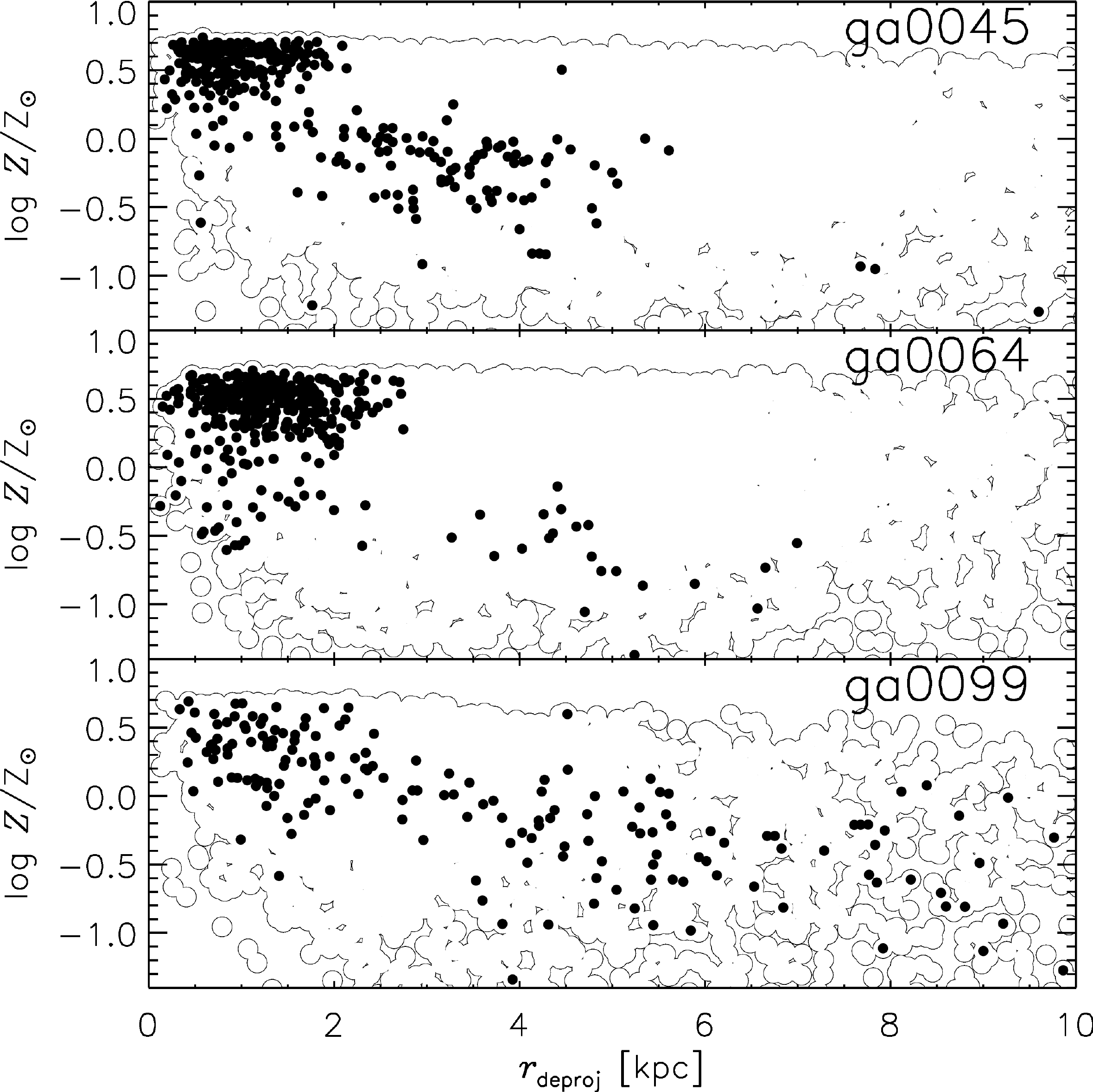}
	\caption{Distribution of stellar metallicity with deprojected (i.e. 3-dimensional) radius for the three galaxies with a CRGD.
	The full distribution is shown in outline; those stars younger than 1 Gyr are shown as solid points.}
	\label{fig:ZsRAge}
\end{figure}

{
The galaxies with CRGDs have stellar metallicities and metallicity gradients that can be $\sim 1\sigma$ from the median trend when $L_V$ weighted.
To understand the reason, in Fig. \ref{fig:ZsRAge} we show the distribution of stellar metallicity with deprojected (i.e. 3-dimensional) radius for these galaxies; the full distribution in shown in outline, and those stars younger than 1 Gyr are shown as full points.
In ga0045 and ga0064, there is a concentration of young, metal-rich stars in the central $\sim 2$ kpc.
These galaxies transitioned from co- to counter-rotating gas approximately 1 Gyr before the present \citepalias{pt18a}, with the bulk of these young stars forming from the last of the co-rotating gas.
These young populations comprise only $\sim4$ per cent of the total stellar mass of these galaxies, but are bright in $L_V$, leading to the $\sim 1\sigma$ effect seen.
Galaxy ga0099 developed a CRGD earlier, about 6 Gyr before the present, and there is no concentration of recent star formation that can have the same effect.
}

The five galaxies have below-average gas metallicity when weighted by mass due to their accretion at late times of near-pristine gas from filaments.
However, when weighted by SFR, three of the galaxies appear more metal-rich than average, which is because star formation is more likely to occur in high-metallicity gas that can cool efficiently.
The gas metallicity gradients are consistent with those of the full sample.
The gas-phase metallicity gradients therefore develop normally, even though their total metallicity is lower.
The values and gradients of [O/Fe]$_{\rm g}$ for these galaxies, and the galaxy population as a whole, are similar regardless of the weighting used.
Two of the galaxies with counter-rotating gas (triangle symbols) have [O/Fe]$_{\rm g}$ gradients away from the average trend, but this is because their [O/Fe]$_{\rm g}$ radial profiles are not well fit by a straight line; see Section \ref{sec:profiles} for more details.

In the bottom-right panel of Fig. \ref{fig:scale}, we show the gas fraction of our simulated galaxies as a function of stellar mass { (gas in any phase throughout the galaxy is included)}.
Three of the five kinematically atypical galaxies lie slightly above the median relation, but they are not the most gas-rich galaxies at a given mass.
Their higher-than-average gas content is likely a consequence of their location within dark matter filaments where gas is readily available, and not a clear indication of the presence of kinematic features.

\subsection{Individual Components}

\begin{table*}
\caption{Stellar metallicity and [O/Fe] for the KDC galaxies ga0074 and ga0091.
	The stellar components that co- and counter-rotate compared to the gas (in the outskirts and the core of the galaxy, respectively), as well as the values for the full galaxy are given.}
	\begin{tabular}[width=\textwidth]{cccccc}
		\multicolumn{2}{c}{\multirow{2}{*}{Galaxy}} & \multicolumn{2}{c}{Mass-Weighted} & \multicolumn{2}{c}{$L_V$-Weighted} \\
		& & $\log Z_*/{\rm Z}_\odot$ & [O/Fe]$_*$ &  $\log Z_*/{\rm Z}_\odot$ & [O/Fe]$_*$ \\
		\hline
		\multirow{3}{*}{ga0074} & co-rotating & $0.17$ & $0.04$ & $0.24$ & $-0.05$ \\
		& counter-rotating & $0.21$ & $0.02$ & $0.24$ & $-0.02$ \\
		& all &  $0.19$ & $0.03$ & $0.24$ & $-0.04$ \\
		\hline
		\multirow{3}{*}{ga0091} & co-rotating & $0.16$ & $0.04$ & $0.10$ & $-0.06$ \\
		& counter-rotating & $0.18$ & $0.03$ & $0.19$ & $-0.01$ \\
		& all &  $0.17$ & $0.04$ & $0.12$ & $-0.04$ \\
	\end{tabular}
\label{tab:kdcz}
\end{table*}

In this section, we separate the components of the CRGDs and KDCs in order to understand the differences in the radial profiles further.
This is only possible with simulations, which have the full kinematic information of individual star particles.
We divide the stars based on their angular momentum, $\mathbf{j}_*$; stars with $\mathbf{j}_*\cdot\mathbf{J}_{\rm *}<0$ make up the KDC, the others orbit  away from the centre of the galaxy, in the same direction as the gas (see \citetalias{pt18a} for more details).
Here, $\mathbf{J}_*$ denotes the total stellar angular momentum of the galaxy, $\mathbf{J}_*=\sum_i\mathbf{j}_{*,i}$.

Table \ref{tab:kdcz} lists the stellar metallicities and [O/Fe] ratios, both mass- and light-weighted, for the components of the two galaxies.
In general, the co-rotating stellar components (in the outskirts of the galaxies) have much lower metallicity than the cores; these stars were almost entirely accreted from the merged galaxy.
Our simulated galaxies follow the mass-metallicity relation \citep{pt16}, and the low metallicities of the co-rotating components reflect the fact that the mergers had small mass ratios ($\sim1/10$).

Although the co-rotating components are in the outskirts, the [O/Fe]$_*$ ratios are only marginally higher when weighted by mass, and are rather lower when weighted by luminosity.
This is due to the longer star formation timescale as discussed for Fig. \ref{fig:maps}.
This is not seen in the observation of NGC 4365 \citep{davies01}, where the [O/Fe]$_*$ is the same between the core and the main body.
Our KDC is a different type and might be found in the next generation of IFU survey with a large field of view such as with Hector (see \citetalias{pt18a} for more discussion).


\section{Conclusions}\label{sec:conc}

Here we presented 2D maps of stellar populations and gas-phase chemical abundances of the simulated galaxies with atypical kinematics shown in \citetalias{pt18a}.
Two of the five galaxies host a minor-merger-originated KDC, and the other three have a CRGD compared to the angular momentum of the stars, caused by delayed filament accretion.

The maps of gas-phase metallicity (Fig. \ref{fig:maps}) showed distinct areas of high and low metallicity separated by a sharp boundary for the galaxies with a CRGD.
The non-linear increase is also seen in the radial profiles (Fig. \ref{fig:zprofile}) but when the gradients are measured, the difference is buried in the large scatter of the gradient-mass relation (Fig. \ref{fig:scale}).
The radial gradients are useful to flag the galaxies formed by major mergers, and the signatures of minor mergers or filament accretion can be detected only with IFU mapping.
Gas-phase metallicity maps are easily obtained from IFU data since the emission lines from ionised gas typically have high signal-to-noise.
We predict that all galaxies that fall from the field into a filament should have gas metallicity maps with a sharp boundary, regardless of the relative orientation of their gas and stellar angular momenta.
Galaxies that form CRGDs via this mechanism are expected to have such metallicity maps, but not all galaxies with such metallicity maps will have a CRGD.
In the galaxies with a CRGD, we also find some non-linear gradients in stellar metallicity; this becomes much clearer when we separate the stars that are formed from the CRGD (Fig. \ref{fig:ZsRAge}), which is possible only in simulations with full 6D kinematics.

For the KDC galaxies, we divide the stars into co- and counter-rotating components using 6D kinematics, and show that the KDC component has lower metallicity than the main component of the galaxy (Table \ref{tab:kdcz}).
We also found that they could be distinguished by their stellar [O/Fe] profiles (Fig. \ref{fig:zprofile}).
At large radii ($r \gtsim 2R_{\rm e}$), the [O/Fe] profiles of these galaxies flattened, and had lower values than the full galaxy population.
This region of the KDC galaxies is made primarily of stars accreted from a minor merger, which formed on a longer timescale than the \emph{in situ} star formation of the primary.
This is a clear prediction for KDCs formed via the mechanism presented in \citetalias{pt18a}, and it is very important to measure [O/Fe] in stelar populations at larger radii ($>2 R_{\rm e}$).
For the same reason, the age gradients tend to be flatter (Figs. \ref{fig:ageprof} and \ref{fig:agegrad}).

None of these kinematically atypical galaxies experiences a major merger, and so there are no significant differences in the scaling relations, except for the size-mass relation.
All of the galaxies with atypical kinematics were found to have lower-than-average values of $R_{\rm e}$, given their masses (Section \ref{sec:scale}).
With major mergers, the stellar distribution becomes kinematically hot, which increases $R_{\rm e}$ \citep[this is also found in the simulations of early-type galaxies in ][]{ck05}.
Major mergers would also destroy the interesting kinematic features of these galaxies, and therefore galaxies with such kinematics should always be compact.

We should note that these features were clear in mass-weighted kinematic maps, but much less discernible in light-weighted maps.
Using stellar populations synthesis models, it is possible to measure both maps with IFU data \citep[e.g.,][]{goddard17}.
In any case, future IFU surveys like Hector \citep{Hector} will need a wide field of view and high sensitivity in order to find these features.

\section*{Acknowledgements}
Parts of this research were supported by the Australian Research Council Centre of Excellence for All Sky Astrophysics in 3 Dimensions (ASTRO 3D), through project number CE170100013.
C.F.~gratefully acknowledges funding provided by the Australian Research Council (Discovery Projects DP150104329 and DP170100603, and Future Fellowship FT180100495) and by the Australia-Germany Joint Research Cooperation Scheme (UA-DAAD).
CK acknowledges support from the UK's Science and Technology Facilities Council (grant ST/R000905/1).
The simulations presented in this work used high performance computing resources provided by the Leibniz Rechenzentrum and the Gauss Centre for Supercomputing (grants pr32lo, pr48pi and GCS Large-scale project 10391), the Partnership for Advanced Computing in Europe (PRACE grant pr89mu), the Australian National Computational Infrastructure (grant ek9), and the Pawsey Supercomputing Centre with funding from the Australian Government and the Government of Western Australia, in the framework of the National Computational Merit Allocation Scheme and the ANU Allocation Scheme.
Finally, we thank V.~Springel for providing {\sc GADGET-3}.


\bibliographystyle{mn2e}
\bibliography{/Users/ptaylor/papers/refs}



\bsp

\label{lastpage}

\end{document}